\documentclass{aa}  
\usepackage{graphicx}
\usepackage{amsmath}
\usepackage{mathtools}
\usepackage{natbib}
\usepackage{txfonts}
\usepackage{amssymb}
\usepackage{verbatim}
\newcommand{\chandra}{{\em Chandra}}
\newcommand{\cdfs}{CDF-S}
\newcommand{\cdfn}{CDF-N}

\begin{document}

\title{What X-ray source counts can tell about large-scale
       matter distribution}

\titlerunning {X-ray source counts}

   \author{A. M. So\l tan  and M. J. Chodorowski  \inst{}       }

   \offprints{A. M. So\l tan}

   \institute{Nicolaus Copernicus Astronomical Center,
              Bartycka 18, 00-716 Warsaw, Poland\\
              \email{soltan@camk.edu.pl, michal@camk.edu.pl} }

   \date{Received ~~ / Accepted }

  \abstract
{Sources generating most of the X-ray background (XRB) are dispersed
over a wide range of redshifts. Thus, statistical characteristics of
the source distribution  carry information on matter
distribution on very large scales.}
{We test the possibility of detecting the variation in the X-ray source
number counts over the celestial sphere.}
{A large number of \chandra\ pointings spread over both galactic
hemispheres are investigated. We searched for all the point-like sources in
the soft band of $0.5 - 2$\,keV  and statistically assessed
the population of sources below the detection threshold.
A homogeneous sample of the number counts at fluxes above
$\sim\!10^{-15}$\,erg\,s$^{-1}$cm$^{-2}$ was constructed for more than
$300$ ACIS fields. The sources were counted within a circular area
of $15$\,arcmin diameter.  The count correlations between overlapping
fields were used to assess the accuracy of the computational methods
used in the analysis.}
{The average number of sources in the investigated sample amounts to
$46$ per field. It is shown
that the source number counts vary between fields at a level
exceeding the fluctuation amplitude expected for the random
(Poissonian) distribution. The excess fluctuations are attributed to
the cosmic variance generated by the large-scale structures. The rms
variations of the source counts due to the cosmic variance within the
$15$\,arcmin circle reach $\sim 8$\,\% of the average number counts.
An amplitude of the potential correlations of the source counts on
angular scales larger than the size of a single pointing remains below the
noise level.}


\maketitle

\section{Introduction}

The largest structures identified in galaxy redshift surveys are
located at distances comparable to the maximum distance at which
structures can be effectively distinguished.  In the CfA redshift
survey beyond the Great Wall not much structure is recognizable at all
(\citealt{geller89}). In the {\it Sloan Digital Sky Survey}
(SDSS)\footnote{http://www.sdss.org/}
the large galaxy filaments are discernible up to roughly $
500$\,Mpc, and the most prominent feature is the Sloan Great Wall
(\citealt{gott05}), found at a distance of $\sim 300$\,Mpc.

Larger structures could potentially be isolated in the quasar part
of the SDSS. Recently \citet{clowes12} report a group of $73$
quasars that span redshifts $1.17 - 1.37$. A total length of this
twisted filament exceeds $1200$\,Mpc. Although \citet{park12} have shown
that the largest structures found in the galaxy distribution are
consistent with the canonical $\Lambda$CDM model, it is plausible
that the quasar alignment could be discordant with this model.

It appears that the question of the large structures could be
addressed using the X-ray surveys.
Most of extragalactic X-ray sources are asociated with various
types of active galactic nuclei (AGN). Because of strong cosmic
evolution and a wide luminosity function, the AGN observed in
the selected flux range are
distributed within a very wide range of redshifts. Equivalently,
the observed source flux is only weakly correlated with the source
distance. Roughly $90$\% of the X-ray background in the
$0.5-2$\,keV energy band is generated by sources brighter than
$3\cdot 10^{-16}$\,erg\,s$^{-1}$cm$^{-2}$ (e.g.
\citealt{moretti03,lehmer12}). A majority of these sources are located at
cosmological distances: $80$\% in the redshift range $0.3 -
2.2$ (\citealt{soltan08}). The comoving radial
distance\footnote{Assuming
``737'' $\Lambda$CDM cosmology, i.e.
$H_o = 70$\,km\,s$^{-1}$\,Mpc$^{-1}$,
$\Omega_m = 0.3$, and $\Omega_\Lambda = 0.7$.}
between
redshifts $0.3$ and $2.2$ exceeds $4200$\,Mpc, thus sources
detected in a moderately deep \chandra\ observation are tracers
of the matter distribution on Gpc scales. One can hardly
expect fluctuations on this scale, but the present
analysis would be capable also of detecting variations on smaller scales,
although with decreasing sensitivity.

In the deepest \chandra\ exposures of $2$\,Ms \cdfn\
\citep{alexander03} and $4$\,Ms  \cdfs\ \hspace{-6pt}
\footnote{http://cxc.harvard.edu/cdo/cdfs.html.}
\citep{xue11}, several
hundred sources are detected. This allows the source distribution to be
investigated along the line of sight, provided the redshifts of
objects are known. However, to investigate large-scale matter
distribution, a transverse dimension of the survey should have a
comparable size to the radial one. In the present paper we explore
the possibility of analyzing the giant structures in the observable
Universe using a large number of the \chandra\ medium depth observations
scattered over the whole sky. Although a single pointing represents
a pencil-like cut through space, a set of several hundred pointings
to some extent act as a uniform all-sky survey.  The main objective
was to select all the pointings suitable for source detection from
the \chandra\ Data Archive, and then to construct an efficient
method of counting the sources in all the fields in a homogeneous way.
A number of sources in each field was used to estimate the amplitude
of the source counts, $N(S)$, at a fixed flux range around
$10^{-15}$\,erg\,s$^{-1}$cm$^{-2}$. A power law parametrization
of the $N(S)$
relationship was applied. Statistical characteristics of the
$N(S)$ amplitude variations are examined. Results, albeit
preliminary, demonstrate the strong and weak points of the method.
The volume of the archived \chandra\ observations is still too
small for a comprehensive study. Nevertheless, the presently
available data reveal the potential advantage of X-ray source counts
in investigations of the large-scale matter distribution.

The organization of the paper is as follows. In the next section,
selection and preparation of the observational data extracted from
the \chandra\ archives is described. Section~\ref{counts} gives
procedures for estimating the amplitude of the number source counts,
$N(S)$. We apply two independent methods to determine the
counts in two separate flux levels. Results of these
calculations, as well as statistical and (possible) systematic
errors are discussed in Sect.~\ref{correlations}. In
Sect.~\ref{variance} we concentrate on the question of
counts fluctuations
that allowed us to estimate the amplitude of the cosmic variance.
The main conclusions are summarized in Sect.~\ref{discussion}.

\section{Selection of observational material \label{selection}}

\begin{figure}
\resizebox{\hsize}{!}{\includegraphics{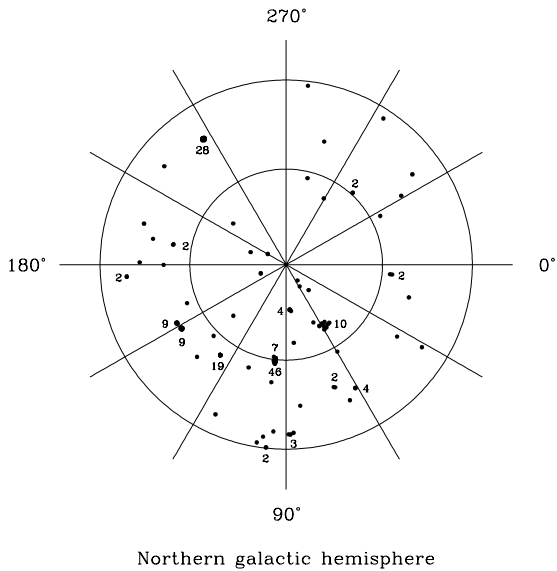}}

\caption{The distribution of $190$ pointings in the northern
galactic hemisphere. Two circles represent galactic latitudes  $b
= 30^\circ$ and $b = 60^\circ$. Each observation covers a circle
of $15$\,arcmin in diameter.  Labels close to some points in the
map denote the actual number of pointings in the area.  Symbol
sizes are not to scale, and even in the crowded areas, the
pointings do not necessarily overlap.}

\label{polar_n}
\end{figure}

Although the total number of the CXO observations now exceeds
$10000$, only a small fraction of pointings is suitable for the
present project. The
number of pointings in ACIS-I configuration longer than
$30$\,ks was equal to $963$ (as for 2012-10-02). Of these $599$ were
located at $|b| > 25$\,deg. Since we are mostly interested in the
fields at high galactic latitudes, without the dominating extended
source, the list of usable data was reduced substantially. In
this step we rejected the pointings at: M87, M33, LMC, SMC, many
Abell clusters (but not all), and the NGC galaxies.
Additionally,  it is desirable to have pointings scattered over the
large area rather than superimposed one onto another or closely
packed. Unfortunately, to some degree the pointings in the CXO
Archive have been accumulated under the opposite criteria.
The archive contains many observations
of nearby galaxies and clusters of galaxies, as well as a large
number of local sources close to the Galactic plane. Also, several
pointings cover the same area (e.g. \chandra\ Deep
Fields).

The \chandra\ Data Archive has been searched for all the ACIS-I
observations at high galactic latitudes with the exposure time
above $30$\,ks. Originally, a selection limit of $|b| > 30^\circ$
has been applied for both hemispheres. Then, only having a few
adequate pointings, we searched for more  observations in a
belt of $25^\circ < b < 30^\circ$ in the southern galactic
hemisphere. Eventually, within this extension just few pointings
satisfied all the selection criteria and were included
in the final set of data.

All the ``interesting''
observations were handled through the standard data processing
pipelines at the \chandra\ X-ray Center\footnote{See the
{\it Chandra Interactive Analysis of Observations} (CIAO) at
http://asc.harvard.edu/ciao.}. Each observation was carefully
inspected for its usefulness for the present investigation.
In particular,
pointings with an xtended source filling substantial fraction
of the field were removed. If the
extended source occupied a relatively small fraction of the field
of view, a section with the source was cut off, and the remaining
data unaffected by the extended emission were used.  Fields
with extremely strong point sources were treated in a similar
way; i.e., only areas affected by linear smearing generated during a
readout were removed.  All the exposures were examined for the
presence of the background flares and only ``good time intervals''
(GTI) were used. Because of a severe deterioration in the image
quality with the increasing off-axis angle, the search for sources was
limited to the circular area of $7.5$\,arcmin radius centered on
chips $0-3$.

\begin{figure}
\resizebox{\hsize}{!}{\includegraphics{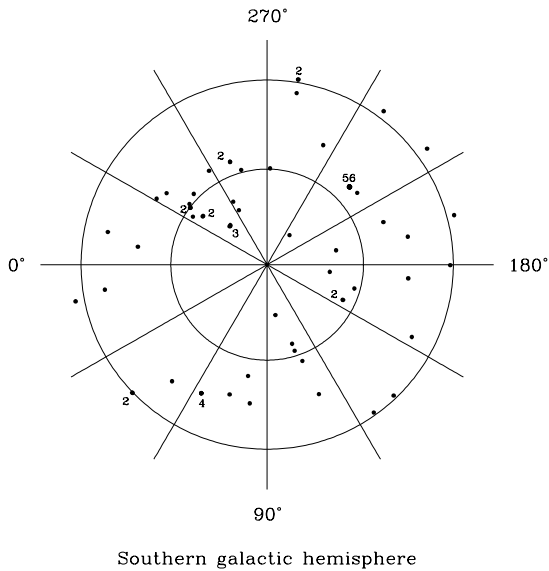}}
\caption{Same as Fig~\ref{polar_n} for $112$ pointings in the
southern galactic hemisphere. A few pointings within
$-30^\circ < b < -28^\circ$ have been included to improve
statistics.}
\label{polar_s}
\end{figure}

Restrictive criteria of the field selection led to elimination
of $~\sim 96$\% of pointings in the CXO Archive. The almost
$400$ observations that remain constitute a sample that is still
numerous enough to make the data processing a painful and
time-consuming operation. Since we are more interested in observations
covering many different directions rather than a single spot in the
sky, we removed some \cdfs\ and \cdfn\ pointings from our sample.
However, we intentionally left several dozen observations in these
areas to facilitate the error analysis (see below). Eventually,
more than $300$ individual pointings were qualified for the
analysis. Their spatial distribution is highly nonrandom.
Concentrations on various angular scales, as well as a strong
asymmetry between galactic hemispheres is present. 

The sky distribution of all the pointings is presented in the polar
equal area projection in Figs.~\ref{polar_n} and \ref{polar_s} for
the northern and southern galactic hemispheres (NGH and SGH, respectively).
Numerals close to some dots denote the numbers of pointings concentrated
in the marked areas. In the NGH, $190$ pointings were selected
for further processing, and $112$ in the SGH.  Symbols in
figures strongly exceed the actual extent in size of the individual
observation. Thus, in some cases where several pointings are
represented by a single dot, the pointings do not always overlap. A
detailed discussion of the statistical properties of the selected
material is presented in Sect.~\ref{correlations}.

The data in a ``canonical'' soft energy band of $0.5-2$\,keV
were used. For various reasons, both the exposure time and
sensitivity are highly variable functions of the position within the
field of view. To reduce the effects of sensitivity variations, low
sensitivity areas have been delineated using the \chandra\
effective exposure map\footnote{An exposure map is generated in a
standard data processing. It is a position-dependent product of
the effective area of the mirror/detector combination and the
effective exposure time [in cm$^2$\,s].}. For each observation, the
maximum amplitude of the exposure map was determined. Pixels with
the exposure map below $75$\% of this maximum value have not
been used.  Although this constraint reduces the original area of
$7.5$\,arcmin circle by roughly $13$\%, it secures for our purpose
an acceptable uniformity of the exposure (\citealt{soltan11}).

\section{`Bright' and `faint' source counts\label{counts}}

Two independent methods have been applied to estimate the number of
discrete point-like sources in a single observation. The first one
was based on a standard procedure of finding prominent
concentrations of counts within a detection cell. Sources isolated
in this way are labeled here as `bright'. A second method, described
in \citet{soltan11}, estimates number of the `faint' sources.
It calibrates deviations of the photon distribution
from the Poissonian one using the nearest neighbor statistics. It
is not able to isolate individual sources, but provides statistical
information on the population of sources generating small numbers of
photons.

To count `bright' sources in a single observation, a circular cell was
slid over the entire investigated area.  The cell size was set to
enclose $85$\% of counts generated by a point source. The radius of the
detection cell varied as a function of the off axis angle according to
the shape of the \chandra\ point spread function (PSF). All the
information on the PSF shape required in the procedure of `bright'
source selection as well as for the `faint' source assessment, was
obtained from {\it mkpsfmap} - a standard tool of the \chandra\
data processing. The search for `bright' sources was performed
in the same manner for all the pointings.  The iterative procedure was
applied. Using the sliding window technique, the strongest
concentration of photons was localized, and it was recognized as the
brightest source in the field of view.  Then, the area surrounding
this source was extracted from the data, and the search for the second
brightest source was performed in the same way. The procedure was
repeated for the consecutive sources until the number of counts
selected as the next source dropped below the threshold adopted
individually for each observation. The threshold counts were selected
at a level high enough to prevent random fluctuation of counts
to be approved as a real source. It was assumed in this step that the
background counts were distributed randomly over the field of view.
The search for sources was terminated when the probability of
random accumulation of counts exceed $10^{-6}$.  Owing to strong
variations in the PSF width, the minimum counts recognized as a
point-like source varies over the field of view, typically between
$4$ and $8$. This range varied a little between observations because
of different levels of the background counts.

Our data span a wide range of exposure times, from $30$\,ks
through more than $170$\,ks. Because of that the pointings differ
in the source detection threshold and in the number of detected
sources. To use the observed number of sources as the estimator
of the source counts per unit solid angle, which is independent of
the exposure time, one needs to adopt an analytic model for the
counts and to relate the observed number of sources in the
individual field to the parameter(s) of the model.

A power law is the obvious choice for the source number counts
per unit solid angle (hereafter per $1$\,sq.\,deg):
\begin{equation}
\frac{dN(S)}{dS} = N_o\, S^a\,,
\label{dNdS}
\end{equation}
with two a priori free parameters, normalization amplitude $N_o$,
and the slope $a$.
However, the limited number of sources, as well as a narrow dynamic
range of fluxes populated by sources in a single pointing, prohibits
any attempts to measure the slope $a$.  Also the data on the
source counts available in the literature for a couple of deep
fields do not provide information on the potential field-to-field
slope variations. In those few cases where the dynamic range
within a single field allows tracking changes of the slope, it
appears that the slope is constant below
$\sim 10^{-14}\,{\rm erg\,s^{-1}\,cm^{-2}}$ down to at least
$10^{-16}\,{\rm erg\,s^{-1}\,cm^{-2}}$
(e.g. \citealt{georgakakis08}).
Therefore, in the present investigation only the normalization is
fitted for each field separately, while a single, universal value
of the slope is assumed.

A several-step procedure was applied to find the best estimate of
$N_o$ using the observed number sources. Because the each
enhancements of counts within the detection cell is considered a
`source', one should
carefully evaluate a contribution of the background counts and
apply the adequate statistical correction. Let $n_s(k)$ be
the number of detected `sources' containing $k$ counts, where
$k\, =\, k_{\rm s}\, +\, k_{\rm b}$
is a sum of the genuine
X-ray photons emitted by a source, $k_{\rm s}$, and the background
counts, $k_{\rm b}$. $N_o$ is related to $n_s(k)$ in the
following way:
\begin{equation}
n_s(k) = A \sum_{k_{\rm b}=0}^{k}\, 
       \left[\, {\rm P}_{\rm bkg}(k_{\rm b})\,
       \int dS\,N_o\, S^a\:
       {\rm P}(k_{\rm s}\,|\,S)\, \right]\,,
\label{ns1}
\end{equation}
where $A$ is the solid angle of the observation (in sq.\,deg),
${\rm P}_{\rm bkg}(k_{\rm b})$ denotes  the probability that
$k_{\rm b}$ background counts would be found in the detection cell,
and ${\rm P}(k_{\rm s}\,|\,S)$ is the probability that a source of
flux $S$ generates $k_{\rm s}$ photons (it is elaborated below).
The probability
of finding $k_{\rm b}$ counts within the detection cell is
given by the Poissonian distribution:
\begin{equation}
{\rm P}_{\rm bkg}(k_{\rm b}) =
         \frac{e^{-\lambda}\,\lambda^{k_{\rm b}}}{k_{\rm b}\!}\,,
\label{poisson}
\end{equation}
where $\lambda$ is the expected number of random counts within
the detection cell. For clarity, in formulae \ref{ns1} and
\ref{poisson}, a fixed detection cell radius and $\lambda$ have
been assumed. In the actual calculations, the right-hand side
of Eq.~\ref{ns1} was integrated over the solid angle $A$.

The sum over the background counts $k_{\rm b}$ in Eq.~\ref{ns1}
generally includes the $k_{\rm s} = 0$ and {1} components.
However, the minimum value of $k$ was set to warrant
${\rm P}_{\rm bkg}(k_{\rm b}=k) < 10^{-6}$, and
those components of the sum are effectively negligible.

It is convenient to use the instrumental count in each
observation as a flux unit. The flux $s$ in ACIS counts
is related to $S$ in physical units by a conversion factor
$\eta$\,: $s=S/\eta$,
with $\eta$ in ${\rm erg\,s^{-1}\,cm^{-2}\,count^{-1}}$
is obtained from the exposure map. The source counts using
the new flux unit are represented by
\begin{equation}
\frac{dn(s)}{ds} = n_o\, s^{\,a}\,,
\label{dnds}
\end{equation}
where
\begin{equation}
 n_o = A\, N_o\,\eta^{\,a+1}\,,
\label{n0}
\end{equation}
and the probability ${\rm P}(k_{\rm s}\,|\,S)$ takes the form:
\begin{equation}
{\rm P}(k_{\rm s}\,|\,s) =
 \frac{e^{-s}\,s^{k_{\rm s}}}{k_{\rm s}\!}\,.
\label{poiss_cnt}
\end{equation}

Equations~\ref{dNdS} and \ref{dnds} with a fixed slope
do not describe the actual counts over the entire flux range.
Substantial fraction of sources is found above the
count slope brake
$S_{\rm b} \approx 8\cdot 10^{-15}\,{\rm erg\,s^{-1}\,cm^{-2}}$
($S_{\rm b}$ is not very accurately determined, see
\citealt{georgakakis08}). Above $S_{\rm b}$ the differential
slope is close to $a_{\rm b} = -2.5$. Incorporating the broken
power law into Eq.~\ref{ns1} and using the instrumental
counts $s$, we finally get
\begin{equation}
\begin{multlined}
n_s(k) = n_o\: A\;\times \\
\times\;\sum_{k_{\rm b}=0}^{k-2}\,
       \left[\, {\rm P}_{\rm bkg}(k_{\rm b})\,
       \frac{\gamma(k_{\rm s}\!+a\!+\!1,s_{\rm b}) +
       s_{\rm b}^{a-a_1}\,\Gamma(k_{\rm s}\!+\!a_{\rm b}\!+\!1,s_{\rm b})}
       {\Gamma(k_{\rm s})}\,\right]\,,
\end{multlined}
\label{ns2}
\end{equation}
where $\gamma(\alpha,x)$ and $\Gamma(\alpha,x)$ are
the lower and upper incomplete gamma functions, respectively.

Equation~\ref{ns2} summed over $k$ gives the expected total number
of `bright' sources. To evaluate $n_o$, the number of actually
detected `bright' sources has been substituted on the left-hand
side of Eq.~\ref{ns2}. Then, the amplitude $N_o$ was obtained using
Eq.~\ref{n0}. Calculating the conversion factor, variations of
soft energy absorption by the cold gas in the Galaxy
were taken into account.
We assumed that the intrinsic source spectra are represented by
a power law with the photon spectral index $\Gamma = -1.4$.
The detected fluxes were corrected for the galactic absorption
using the appropriate $N_H$ data.
We used the NASA HEASARC tool\footnote
{http://heasarc.gsfc.nasa.gov/cgi-bin/Tools/w3nh/w3nh.pl}
based on \citet{kalberla05}.
The lowest $N_H$ in the sample is $6\cdot 10^{19}$,
and highest is $1.6\cdot 10^{22}\,{\rm cm}^-2$. 
For the $0.5-2$\,keV energy band and the assumed spectrum,
this $N_H$ range introduces variations of the average photon energy
between $1.022$ and $1.154$, and a reduction of detected counts by
a factor $0.984$ through $0.695$.
The statistical properties of the $N_o$ distribution
in our sample are discussed in the next section in conjunction
with the analogous distribution obtained for `faint' sources.

The population of sources below the formal source detection
threshold, labeled as `faint', is investigated using a
statistical approach. Here only the basics of the method are
presented. All the details and potential applications were
discussed in a separate paper (\citealt{soltan11}).

After the removal of all the `bright' sources, the remaining counts
are a mix of the non-X-ray events, truly diffuse X-ray background,
and counts generated by a population of faint sources. Only this last
constituent is responsible for fluctuations of the count
distribution characteristic for point sources. One can express the
total number of events, $n_t$, within the investigated area 
in the following form:

\begin{equation}
n_t\, =\, n_1 + n_2 + ... + n_{k_{\rm max}}\,,
\label{nt}
\end{equation}
where $n_1$ represents counts that are distributed randomly, $n_2$ -
counts produced by sources contributing exactly two counts each
(`2-photon sources'),
$n_3$ - counts by `3-photon sources', and so on up to
$n_{k_{\rm max}}$ - counts due to the brightest sources left
in the field (i.e., the brightest among the `faint' ones).

Counts distributed randomly constitute a composite
collection of events that also includes the
weakest discrete sources contributing single photons.
Photons coming from sources producing $2\le k \le k_{\rm max}$
counts create local enhancement of a unique shape determined
by the telescope PSF.
Statistical characteristics of these variations are efficiently
quantified using the nearest neighbor
statistics (NNST) of the count distribution. The probability that
a randomly chosen event has no neighbors within a distance $r$, is
\begin{equation}
P(r) = p_1\, P(r\,|\,1)\, +\, p_2\, P(r\,|\,2)\, + ... +
        p_{k_{\rm max}} P(r\,|\,k_{\rm max})\,,
\label{pr}
\end{equation}
where $p_1$ is a probability that this event is not related to any
source generating two or more counts, $p_k$ for $k = 2, ...,
k_{\rm max}$ is the probability that this event is produced by
the $k$-photon source,
and $P(r\,|\,k)$ is the conditional probability that there are
no other counts within $r$ provided the selected event belongs
to the $k$-photon source. Under the reasonable assumptions, for
$k\ge 2$
\begin{equation}
P(r\,|\,k) = P(r\,|\,1)\cdot {\cal P}(r\,|\,k)\,,
\label{prk}
\end{equation}
where ${\cal P}(r\,|\,k)$ is the probability that within $r$
there are no other counts from the same source. This quantity
is uniquely defined by the PSF. Taking $n_k = k\cdot n_s(k)$ and
$p_k = n_k/n_t$ and combining Eqs.~\ref{pr}, \ref{prk}, \ref{nt},
and \ref{dnds} we get
\begin{equation}
\frac{n_o}{n_t}\, P(r\,|\,1)\,
     \sum_{k=2}^{k_{\rm max}}\, \frac{\Gamma(k+a+1)}{\Gamma(k)}\,
     [1 - {\cal P}(r\,|\,k)]\, =\, P(r\,|\,1) - P(r)\,.
\label{nnst} 
\end{equation}

In the relevant flux range the source counts are adequately
represented by a single power law. Since the maximum number of
photons, $k_{\rm max}$, is defined by the threshold for the `bright'
sources, the actual value of $k_{\rm max}$ is a function of a
position within the field of view; in most of the pointings
$3\le k_{\rm max}\le 7$.

Equation~\ref{nnst} allows us to determine $n_o$, provided the
relevant probability distributions are known. One can estimate the
probability $P(r)$ for each observation by measuring the distance to
the nearest neighbor for each photon in the field. Similarly,
$P(r\,|\,1)$ is given by the distribution of distances of the
randomly distributed points to the nearest photon.
The PSF related probability distributions, ${\cal P}(r\,|\,k)$,
were determined using the CIAO tool {\it mkpsfmap}. The distribution
of photons produced by a large number of sources were generated by
means of the Monte Carlo method, and the relevant probabilities
were determined as a function of position and number of photons.

The efficiency of the NNST method depends strongly on the PSF width,
and in the present investigation it provides interesting results
only for small off-axis angles. Because statistical uncertainties
increase rapidly at larger distances from the optical axis, we limited
the NNST calculation to the central region of $5$\, arcmin diameter.

The best estimate of $n_o$ was determined in the following way.  All
the probability distributions were calculated for several
(usually more than $20$) distances $r$, which uniformly covered the
domains of distributions $P(r\,|\,1)$ and $P(r)$.  Then, the
cumulative probability distributions in Eq.~\ref{nnst} were
replaced by the corresponding probability densities; e.g.,
$\Delta P(r) = P(r) - P(r + \Delta r)$ for the consecutive
distances $r$.
In all the calculations, $\Delta r = 0.246$\,arcsec ($\equiv 1/2$ of
the instrumental pixel) was used. Thus, a set of independent
equations covering all the observed separations was
constructed, and the best fit value of $n_o$ was determined using
the least square method. Finally, the count amplitude $N_o$ was
obtained using Eq.~\ref{n0} as for the 'bright' sources.

\section{Count correlations \label{correlations}}
 
The amplitudes of the number source counts determined for `bright'
and `faint' sources (hereafter denoted $N_o[b]$ and $N_o[f]$,
respectively) define the number source counts over adjoining, but
different flux levels. Therefore, the median flux of `bright' and
`faint' sources in each observation is not widely separated, and the
$N_o[b]$ and $N_o[f]$ averaged over the sky should be similar.
Nevertheless, one can expect systematic differences between both
estimates, if the count slope adapted in the calculations, $a$, did
not match the slope actually observed in the data. The differential
slope between $10^{-16}$ and $10^{-14}$\,erg\,s$^{-1}$cm$^{-2}$ in
the deep \chandra\ fields is well fitted with the slope of $-1.58$
(\citealt{georgakakis08,soltan11}). Using $a = - 1.58$ in the present
calculations, the average amplitude of the `faint' source counts is
larger by $\sim\!5$\% then that for the 'bright' sources. Both
amplitudes are practically identical for $a = 1.60$.  To preserve
internal consistency of the investigation, the latter slope was
used in subsequent calculations. Although this is steeper by
$0.02$ than the best fit by \citet{georgakakis08}, it stays within
their $1\sigma$ limits. One should note that the present
investigation is not suitable for assessing the count slope either within
the individual field or the average over the sky.  This is because
most systematic errors involved in the estimates of the number of
`bright' sources are independent of the errors affecting the
`faint' source assessment. The source counts amplitudes, $N_o[b]$
and $N_o[f]$ are functionally dependent on the slope $a$.
Therefore, a unique slope is required to balance both amplitudes. In
fact, modification of the slope by just $0.02$ demonstrates that the
systematics do not play a significant role in the analysis. One
should also stress that all the conclusions of the paper are not
affected by a selection of the particular value of the slope.

An apparent consistency of the average $N_o[b]$ and $N_o[f]$
estimates does not exclude local inhomogeneities of the source
distribution in the individual pointing.  One should note that the
amplitudes $N_o$ for `bright' and `faint' sources are obtained on
using the disconnected classes of sources; i.e., none source was
qualified as `bright' and 'faint' at the same time. Thus, for the
random source distribution, the source counts should be
uncorrelated.

\begin{figure}
\resizebox{\hsize}{!}{\includegraphics{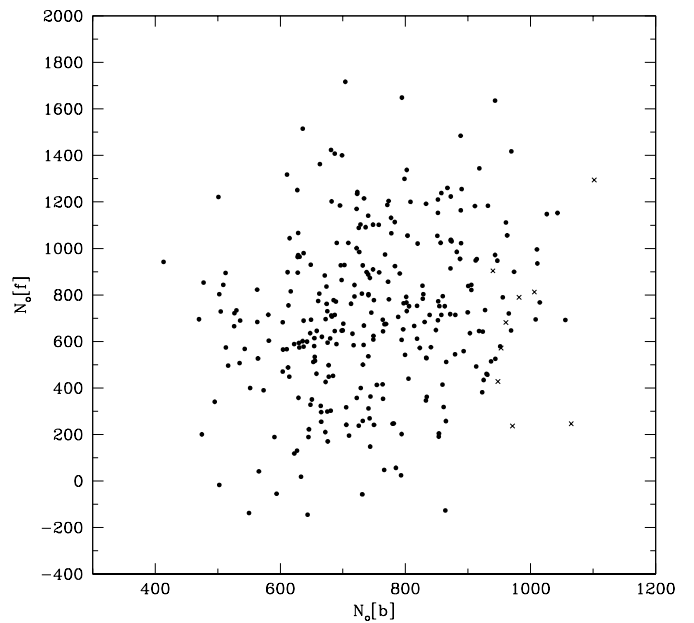}}
\caption{Amplitudes $N_o$ of the source number counts for flux $S$
in $10^{-15}$\,erg\,s$^{-1}$cm$^{-2}$ (see Eq.~\ref{dNdS}) for
the `bright' and `faint' sources in the sample of $302$ pointings;
$293$ pointings with the rms uncertainty of $N_o[b]$
smaller than $160$ per sq.\,deg are shown with dots, $9$ pointings
with larger rms - with crosses.} 
\label{nobnof}
\end{figure}

Apart from the systematic errors, the $N_o[b]$ and $N_o[f]$
amplitudes are subject to large statistical uncertainties that
generate substantial scatter.  The average number of `bright'
sources per field ${\overline n_s} = 46.0$. Thus, the Poissonian
scatter alone introduces $\sim\!15$\% rms fluctuations to the
present estimate of $N_o[b]$. In reality, still larger variations
of $N_o[b]$ are expected due to  cosmic variance (source
clustering). The $N_o$
estimates for the `faint' sources are afflicted by even larger
errors, because the estimate of the count amplitudes is derived
indirectly. Two factors contribute most to the final
uncertainties. First, the amplitude estimate is proportional to the
difference between two probability distributions (see
Eq.~\ref{nnst}), and both distributions are subject to substantial
statistical scatter. Second, the `faint' sources' contribution to
the total events is small, typically $2 - 6$\%, while all the
remaining counts are distributed randomly. Therefore the net
signal, represented by the right-hand side in Eq.~\ref{nnst}, is
also very modest in comparison to the noise that is dominated by
the random scatter of the $P(r\,|\,1)$ distribution.

In Fig.~\ref{nobnof} the `bright' and `faint' source amplitudes $N_o$
as defined in Eq.~\ref{dNdS}\footnote{In the calculations broken
power law and Eq.~\ref{ns2} have been used.}, where $S$ is in
$10^{-15}$\,erg\,s$^{-1}$\,cm$^{-2}$, are plotted for the full sample
of $302$ pointings. The rms scatter of $N_o[b]$ is slightly
larger than what is expected from the Poissonian fluctuations of the
`bright' source numbers (see below for the detailed discussion),
while the amplitude of the $N_o[f]$ variations is much higher, and
it is clearly dominated by the statistical noise.
The rms for 'faint' sources amounts to $44$\% of the average
amplitude of $N_o[f]$. Despite the scatter, the correlation of the
`bright' and `faint' amplitudes is statistically significant. The
correlation coefficient for $302$ data points $\rho = 0.2003$.  The
non-directional probability that the data are drawn from the
uncorrelated population amounts to $0.00046$ (or $0.00023$ for
the directional case).  To minimize statistical noise, the
uncertainties of the individual $N_o$ estimates are carefully examined
(see below). Due to a wide range of exposure times in the processed
observations, these uncertainties strongly differ.
In order to check the influence of the poor quality estimates on the
subsequent statistical analysis, pointings with the largest
uncertainties of $N_o[b]$ have been identified. A removal of
pointings that suffer from the highest uncertainties does not
strengthen the $[b] - [f]$ correlation greatly. Nevertheless,
if nine pointings with $\sigma_{\rm o}(N_o[b]) > 160$ per sq.\,deg
are eliminated, the correlation coefficients in the sample of
$293$ pointings rises to $\rho = 0.2205$. The corresponding
directional probability for the uncorrelated general population
in this case drops to $7\cdot 10^{-5}$.

Although the $N_o[f]$ estimates are strongly affected by the
statistical noise at the level of the photon distribution within
the each pointing, the correlation between $N_o[f]$ and
$N_o[b]$ in the whole sample favors the calculations of a single
amplitude $N_o$ derived jointly from the `bright' and
`faint' sources.  To combine the $N_o[b]$ and $N_o[f]$
estimates into a single `best' estimate of the source number counts
amplitude for each pointing, an adequate error estimate and data
weighting for each measurement are essential. The errors for the
`bright` source counts were calculated using the Poissonian
noise, since this effect dominates. In the first step it was
assumed that the underlying population of sources is the same for
all the pointings; i.\,e., the cosmic variance effects were
neglected. Next, the expected number of the observed
`bright' sources was determined for each pointing separately,
taking the exposure time into account. Finally, the Poissonian
distribution around the expected number of sources was used to
compute the corresponding dispersion $\sigma_{\rm o}(N_o[b])$
individually for each pointing. The rms uncertainties are
contained between $64$ and $182$ per sq.deg with the average in
the whole sample of $112$ per sq.\,deg. The rms scatter of $N_o[b]$ 
amounts to $127$, which exceeds the Poissonian estimate almost by
$14$\%.

The $N_o[f]$ uncertainties result from a complex combination of the
Poisson-like fluctuations of the number of faint sources,
contribution of background counts, and a stochastic character of
the photon distribution on angular scales comparable to the width
of the PSF. In effect, the variance of $N_o[f]$ cannot be easily
derived from theoretical considerations, but instead it is
estimated directly from the observed distribution. It was evident
that the cosmic variance contributed marginally to the observed
$N_o[f]$ scatter. Also, no clear correlation of the variance with
the exposure time (or the total number of events) was found. 
Therefore, equal uncertainties were assigned to all the
observations. As an error of the individual $N_o[f]$ estimate,
$\sigma_{\rm o}(N_o[f])$, the
rms scatter in the whole sample had been assumed: $\sigma(N_o[f]) =
337$ per sq.\,deg. 

Our final best estimate of $N_{oi}$ of the $i$-th observations
is a weighted mean of $N_o[b]_i$ and $N_o[f]_i$:
\begin{equation}
N_{oi} = \alpha_i\cdot N_o[b]_i + \beta\cdot N_o[f]_i\,,
\label{weighted_n_o}
\end{equation}
with weights $\alpha_i$ and $\beta$ inversely proportional to the
squares of the corresponding rms uncertainties. The resulting
uncertainties of $N_o$, $\sigma_{\rm o}(N_o)$, are spread between
$63$ and $161$ per sq.\,deg.

The drastic unevenness  of the pointing distribution in the
celestial sphere presents a problem for our statistical analysis.
Estimates of the average count amplitude, as well as the count
correlation between different pointings, should take into account
fact that a large number of pointings cover the same sky area.
Obviously, the source counts based on pointings that cover the
overlapping areas of the sky are related. For the detached
(nonoverlapping) observations, the correlations would indicate
structures in the source distribution on the corresponding angular
scale. Since our sample contains a large number of overlapping
pointings, the spurious count correlations are also present at
larger angular scales than the field of view of a single pointing.
This is because the overlapping observations are dominated by a few
groups centered on \chandra\ deep fields. Some wide angular bins
that happen to include the separation between these fields contain
pairs predominantly drawn from those heavily observed areas. To
account for this bias, we assign weights to all the pointings.  As a
result of the application of weights, cosmic signals from all the
directions covered by the present observations contribute uniformly
to the investigated correlations. The weight of the $i$-th pointing
is defined by the formula

\begin{equation}
w_i\; =\, \Omega_i\; /\int_{\Omega_i} m_{\rm pnt}\,d\omega\,,
\label{weights}
\end{equation}
where $\Omega_i$ denotes the solid angle of the observation, and
$m_{\rm pnt}$ is the number of pointings that cover the area
$d\omega$. Thus, for all the detached pointings, $w = 1$ since
$m_{\rm pnt} = 1$ within the entire field of view, while for the
coinciding pointings, the weights are reduced proportionally to the
overlapped area and the number of the involved pointings. Unless
otherwise stated, these weights are applied in the calculations
below.

The average number counts amplitude $\overline{N}_o = 743$ per
sq.\,deg. with the rms scatter $\sigma(N_o) = 125$ per sq.\,deg.
The flux limits that define `bright' and `faint' sources depend on
the exposure time, but the flux separating both source classes is
always close to $10^{-15}$\,erg\,s$^{-1}$\,cm$^{-2}$. Therefore the
present $\overline{N}_o$ estimate characterizes the source counts
just around this flux level.

To compare the present result with the source number counts based on
several deep \chandra\ fields, we adopt the count parametrization by
\citet{georgakakis08}. Their data converted to the units used here
give $dN/dS = 574$ per sq.\,deg at
$S=10^{-15}$\,erg\,s$^{-1}$\,cm$^{-2}$; i.e., our figure is $29$\%
higher. Although this difference could partially be attributed to
various systematic effects inherent in the present calculations, one
should note that our amplitude is the average of a large number of
pointings distributed over the whole sky, while the \chandra\ Deep
Fields only cover a few selected areas. Applying the present method to the
selection of $47$ observations in the \cdfs\, we get the average
amplitude $N_o = 641 \pm 8$, per sq.\,deg, which exceeds the
\citet{georgakakis08} result by $12$\%. The uncertainty of the
average is based on the rms scatter of $56$ per sq.\,deg in those
$47$ pointings. Although all these data are localized at \cdfs, the
fields cover slightly different areas. Therefore variations of $N_o$
not only represent the statistical noise generated by our
procedures, but also include small contribution from the actual
fluctuations in the number of sources.

In the \cdfn\ the present calculations give $dN/dS = 797 \pm
17$ per sq.\,deg. This amplitude differs from the \cdfs\ one by
$24$\% and it is consistent with results by \citet{tozzi01} for
\cdfs, and \cdfn\ by \citet{brandt01}.

\begin{figure}
\resizebox{\hsize}{!}{\includegraphics{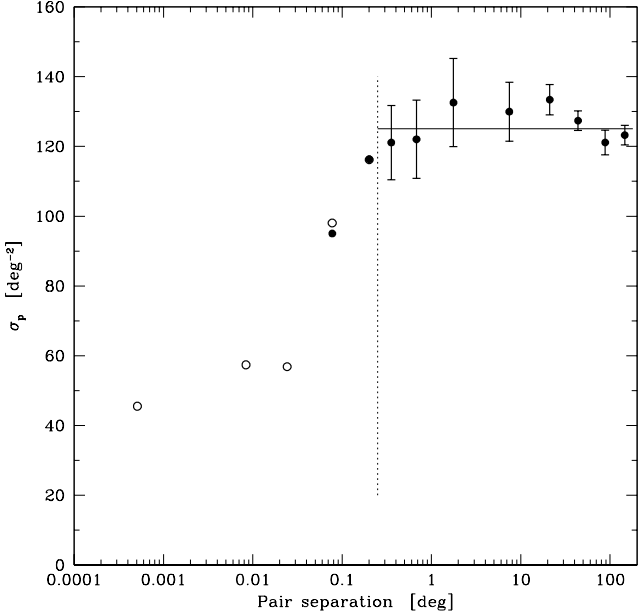}}
\caption{The distribution of the `pair rms', i.e. rms difference
between the counts amplitudes in pairs of observations as a function
of the pair separation. Results obtained with and without weights
are shown with full and open circles, respectively (see text for
detailed explanation). The dotted line separates the overlapping
pointings from the detached ones. The lowest separation bin contains
pairs closer than $0.1$\,arcmin.}
\label{pair_rms}
\end{figure}

To assess the uncertainties proper to the method, and cleared of
the contribution generated by the cosmic variance, all $302$
observations (including overlapping)  have been used in the
following way. For all the possible pointing pairs, a difference
between amplitudes $N_o$ was obtained: $\Delta_{ij}=N_o(i)-N_o(j)$,
for $i, j = 1, ...,\, 302$, and $i\ne j$. The pairs were divided in
several bins according to the pair angular separation Then, for the
each separation bin the rms of $\Delta_{ij}$ was calculated:

\begin{equation}
\sigma_p^2 = \frac{1}{N_{ij}}\,
             \frac{1}{2}\sum_{i\ne j} \Delta_{ij}^2\,,
\label{sigma_p}
\end{equation}
where $N_{ij}$ denotes the number of pointing pairs in the given
separation bin, and `2' in the denominator is introduced to
normalize $\sigma_p$ to the rms of $N_o$. For perfectly overlapping
pointings, the `pair rms' $\sigma_p$ represents the scatter
intrinsic to the method, while for widely separated fields,
$\sigma_p$ is a quadratic sum of two components: noise produced
by Poissonian scatter of the source counts detected in each pointing
and the cosmic variance. If the observations partially overlap,
intermediate amplitudes of $\sigma_p$ are expected.  In
Fig.~\ref{pair_rms} the distribution of $\sigma_p$ for several bins
is shown. The abscissa positions give the average pair separation in
the bin.
One sigma error bars for detached pairs were calculated using the
bootstrap Monte Carlo method. A large number ($10000$) of simulated
sets of observations were generated.  In the simulated data the
original sky coordinates were used, while the amplitudes $N_o$
were drawn at random from the observed $N_o$ distribution. Then,
using Eq.~\ref{sigma_p} the pair rms $\sigma_p$ were calculated
for each simulated data set. The rms scatter of the $\sigma_p$ was
taken as the uncertainty of the actually observed signal.

Weights are superfluous for almost completely overlapping pointings,
since the rms mainly represents  the method errors, while weights adequately
measure the contribution due to the cosmic signal for the detached pairs.
To calculate the pair rms in that latter case  we use the formula:
\begin{equation}
\sigma_p^2 = \frac{1}{2}\left.\sum_{i\ne j} w_i\, w_j\, \Delta_{ij}^2
             \;\middle/\,\sum_{i\ne j} w_i\, w_j\right..
\label{wgt_sigma_p}
\end{equation}

At the smallest separations, the amplitude $\sigma_p = 41.6$\,per
sq.\,deg, and for the `bright' sources $\sigma_p[b] = 37.5$\,per
sq.\,deg.  A substantial decline in the pair rms for overlapping
pointings clearly indicates that the observed fluctuations are
dominated by the actual variations of the cosmic signal while
shortcomings of all the procedures and approximations engaged in the
calculations do not play a significant role. One should also note
that the rms scatter attributed here to the present method does not
affect the Poissonian character of the distribution of the number of
sources detected in a single observation. Thus, it is legitimate to
assume that the statistical properties of the $N_o$ distribution in
the sample are described by the cosmic variance and the Poissonian
scatter.

\section{Cosmic variance} 
\label{variance}

No obvious variations of the pair rms is present for the
nonoverlapping pointings averaged over all the data. It indicates
that any potential, strong fluctuations of the counts amplitude have
angular scales that typically do not exceed a size of the individual
pointing. However, on smaller angular scales, fluctuations
exceeding the Poisson noise are present in our sample.

Assuming that the cosmic variance and the Poissonian noise add
in squares to produce the observed scatter of $N_o[b]$, the rms of
the cosmic variance $\sigma_{\rm CV} \approx 60$, or $8.1$\% of
the average $N_o[b]$ signal. This figure represents the mean
variance within a circle of $15$\,arcmin diameter. One should expect
that the cosmic variance amplitude also depends on the source
redshifts.  However, over a wide range of X-ray fluxes, the average
source redshift depends very weakly on the flux threshold (e.g.
\citealt{soltan08}). Thus, even for a wide span of exposure times,
the redshift distributions in the sample are statistically similar,
and the $\sigma_{\rm CV}$ amplitude is representative of all
our data.

The statistically significant correlation between the `bright' and
`faint' source counts also affirms the role of the cosmic variance.
The correlation alters the rms fluctuations of the combined counts
amplitude $N_o$. To test this effect on the distribution of $N_{o}$,
we generated a large number of simulated sets of count amplitudes
using Eq.~\ref{weighted_n_o}. In the simulated data we connected the
actual `bright' source amplitude of the $i$-th observation,
$N_o[b]_i$, with the $N_o[f]$ randomly drawn from the sample.  It
was found that the rms fluctuations of the $N_o$ in the simulated
data are in $99$\% cases smaller than the fluctuations in the real
data, which confirms the $N_o[b] - N_o[f]$ correlation revealed using
the Pearson test.

\section{Discussion\label{discussion}}

Variations of the source number counts were analyzed using the
\chandra\ observations distributed in both galactic hemispheres.
We applied two independent methods determining the source
counts in the two adjacent flux levels. Then, both amplitudes for
each pointing were reduced to the standard flux of
$10^{-15}$\,erg\,s$^{-1}$cm$^{-2}$.
The amplitudes represent separate species of sources. Thus,
for the Poissonian distribution of sources both amplitudes are
expected to be statistically uncorrelated.

The dominating contribution in the observed field-to-field
variations comes from the Poissonian noise induced by the relatively
small average number of sources per field. This effect seriously
confused a search for the potential correlations in the $N_o$
distribution on large angular scales. Scarce observational data,
particularly in the south galactic hemisphere, prevent us from
drawing conclusions at a high confidence level. The pointings are
distributed very inhomogeneously in the celestial sphere on the
large, as well as on small angular scales. Due to these deficiencies,
the picture emerging from the present investigation is ambiguous
to some extent. 

On large angular scales, the fluctuations of the counts
amplitude averaged over the whole sample of $302$ observations do
not exhibit distinct irregularities that could be symptomatic for
the existence of the large angular scale structure of the XRB.
However, the statistically significant correlation between
`bright' and `faint' sources within individual pointings is clearly
visible, implying a nonrandom source distribution on adequately
small scales.  The present data are insufficient to measure
the angular range of these correlations precisely. A sharp rise of
$\sigma_p$ with increasing separation for the overlapping pointings
(between $3$ and $15$\,arcmin; see Fig.~\ref{pair_rms}) indicates
that the angular size of fluctuations is most likely smaller than
the extent of a single pointing. A substantially larger collection
of closely separated observations is needed to determine this
question.

Despite the apparent absence of large angular scale features in
the $\sigma_p$ distribution, some differences between the data in
both galactic hemispheres seem to be present. Unfortunately, the
available observational material is inhomogeneous, which makes the
comparison between N and S quite problematic. In particular, the
number of pointings in the northern and southern hemispheres are
$190$ and $112$, respectively. An asymmetry is even stronger in the
total area covered by the pointings: in the north it is twice as big
as in the south.

We notice that the fluctuations of $N_o[b]$ in the north are
noticeable larger than in the south. Since the average numbers of
detected `bright' sources per field in both areas are very similar,
the Poissonian contribution to the rms of $N_o[b]$ are also equal and
amount to $112$ per sq.\,deg. In effect, the whole difference is
attributed to the cosmic variance, which in the north exceeds $68$
per sq.deg, while in the south $\sigma_{\rm CV} = 41$ per sq.\,deg.
Also the $N_o[b] - N_o[f]$ correlations are substantially different
in both hemispheres.  In the north the correlation is statistically
significant at the confidence level of $99.98$\% (using the
directional probability), while the data in the south are consistent
with no correlation. It is complicated to interpret these
discrepancies. Although some statistical characteristics of count
amplitudes in both hemispheres look essentially different at first 
glance, it is most likely that these ``unusual'' features are
artifacts that should be attributed to the specific properties of
this particular set of observations, rather than to the genuine
large-scale variations of the source distribution.

A large number of the new pointings would improve statistical
significance of the investigations. Unfortunately, one cannot expect
a rapid inflow of the new \chandra\ observations. Therefore, a more
promising way to proceed would be to search for the correlation
between the existing \chandra\ observations and the data in other
available X-ray surveys.  In particular, the relationship between
the counts amplitude determined here and the RASS maps should be
investigated. 

\begin{acknowledgements}
We are grateful to the anonymous
referee for careful reading and for for remarks and propositions
for improving the paper that resulted in the substantial and -- in
some points -- essential changes in our work.
We thank all the people who created the \chandra\ Interactive
Analysis of Observations software for producing a user-friendly
environment.  This work has been partially supported by the Polish
NCN grant 2011/01/B/ST9/06023.
\end{acknowledgements}

\end{document}